\documentclass[11pt,twoside]{article}

\usepackage{asp2006}
\usepackage{epsf}
\usepackage{lscape}

\begin{document}
\title{Modeling the slow solar wind during the solar minimum}   
\author{Leon Ofman and Maxim Kramar}   
\affil{The Catholic University of America, NASA-GSFC, Code 671, Greenbelt, MD 20771, USA}    

\begin{abstract} 
During the solar minimum STEREO observations show that the
three-dimensional structure of the solar corona can be described
well by a tilted bi-polar magnetic configuration. The slow solar
wind is modeled using three-fluid model that includes heavy ions,
such as He II and O VI. The model is initialized with dipole
magnetic field and spherically symmetric density. The resulting
steady state non-potential and non-uniform streamer configuration
calculated with this model is compared to STEREO observations of the
streamer density structure. SOHO/UVCS observations are used to
compare the O VI emission to model results. We discuss the unique
properties of the solar wind produced in this configuration.
\end{abstract}


\vspace{-0.7cm}
\section{Introduction and Motivation}
Observations show that the slow solar wind is associated with the
streamer belt during solar minimum, and the magnetic configuration
near solar minimum is dominated by a tilted dipole. Past UVCS
observations show that the emission structure and abundances of O~VI
and other heavy ions in streamers are different than the properties
of Ly~$\alpha$ emission \citep{Kohl1997,Raymond1997}.

\begin{figure}[h]
\epsscale{0.9}\plotone{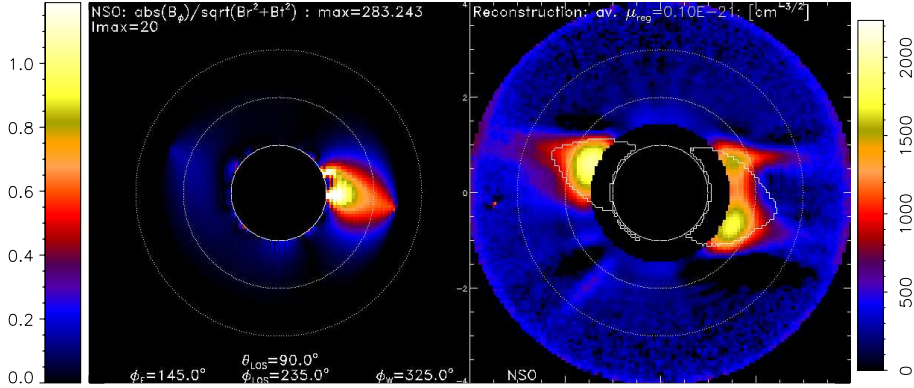} \caption{Right panel:
Cross-sections of the reconstructed electron density in square root
scale on a plane perpendicular to a LOS with Carrington longitude of
$235^\circ$ and colatitude of $90^\circ$. The white contour lines
are the boundaries between closed and opened magnetic field lines
for a PFSS model with source surface at $2.5 R_\odot$ for CR 2066.
Left panel: Cross-sections of the value
$|\vec{B}_\phi|/\sqrt{\vec{B}_r^2+\vec{B}_\theta^2}$ in the PFSS
model on the same plane showing that this quantity is small.}
\label{Fig_vert}
\end{figure}

As evident from recent NSO and STEREO 3D reconstruction of electron
density, and magnetic field using Potential Field Source Surface
Model (PFSS) the streamer higher density region is confined to the
central part of the tilted dipole with nearly symmetric structure
with respect to the tilt plane \citep{Kramar2009}.
Figure~\ref{Fig_vert} shows and observational example that justifies
the use of 2.5 MHD with dipole field. Global 3D MHD models of the
solar wind can capture the overall configuration of the streamers
\citep{Mikic1999}. However, these models can not describe the
different properties of the various ions.

Recently, \citet{Ofman2000,Ofman2004a} developed the first
three-fluid model of a coronal streamer with O$^{5+}$, and He$^{++}$
ions included self-consistently as a third fluid in addition to
electrons and protons using quadrupole initial magnetic field. The
model was able to reproduce the main observational properties of the
heavy ions. Here we present an extensions of the three-fluid 2.5D
model of streamers near solar minimum with heavy ions, initiated
with dipole magnetic field appropriate for solar minimum, that
evolves into the streamer structure in the quasi-steady state. We
find that the model reproduces the properties of the slow solar wind
protons and of the heavy ions.

\section{Multi-fluid model}

We solve the normalized multi-fluid equations for $V\ll c$, with
gravity, resistivity, viscosity, and Coulomb friction, neglecting
electron inertia, assuming quasi-neutrality. The equations can be
found in \citet{Ofman2000,Ofman2004b}. In \citet{Ofman2000} the
heavy ions were isothermal. In the present study we use more
realistic Polytropic model with energy coupling term between the
ions \citep{Bra65} with the following energy equation:
\begin{eqnarray}
%
%
%
%
&&\frac{\partial T_k}{\partial t}=-(\gamma _k-1) T_k
\nabla\cdot\vec{V}_k - \vec{V}_k\cdot\nabla T_k +
C_{kjl}\label{T:eq},
\end{eqnarray}
where $k=e,\ p, i$, and $\gamma _k=1.05$ is the polytropic index.
The energy exchange terms between the species in
equation~(\ref{T:eq}) are given in \citet{Ofman2004b}.

In order to reproduce solar minimum conditions we initialized the
3-fluid model with a dipole field (see Figure~\ref{field_3f:fig}a).
The density was initially uniform in $\theta$, the radial dependence
of proton density was given by spherically symmetruc Parker's solar
wind solution. The heavy ion abundance was initially a constant
fraction of the proton density everywhere (i.e., 0.05 for He$^{++}$
and $8\times10^{-4}$ for O$^{5+}$ ions). These values evolved
self-consistently as the slow solar wind streamer has formed in the
qausi-steady state.

\section{Numerical Results}
\begin{figure}[h]
\epsscale{0.8}\plotone{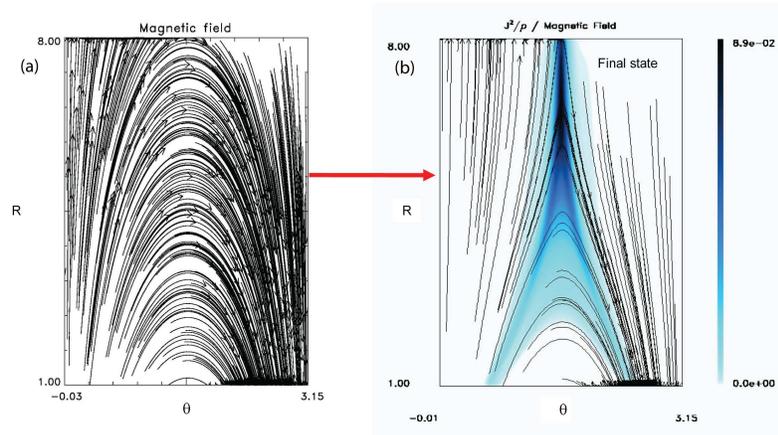}\vspace{-0.5cm} \caption{The
transformation of (a) dipolar field into (b) streamer obtained with
the 3-fluid model is show. The current-sheet is shown in blue scale.
} \label{field_3f:fig}\end{figure}

\begin{figure}[h]
\epsscale{0.8}\plotone{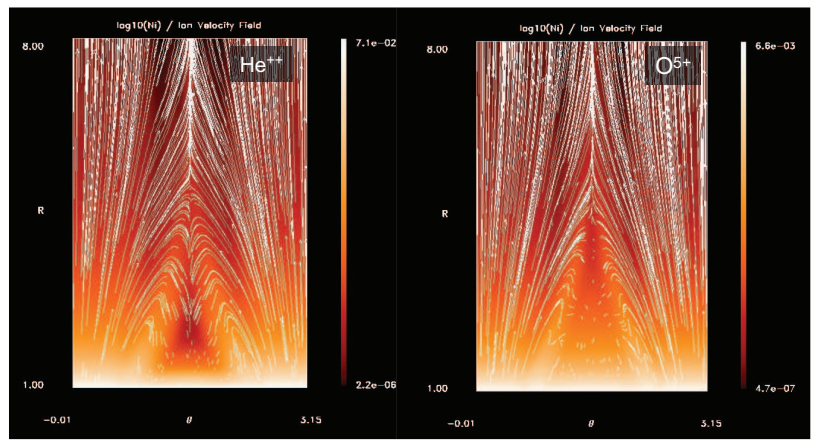} \caption{The density
of the ions in the slow solar wind  streamer overlaid with the ions'
outflow speed. } \label{str_3f:fig}\end{figure}

\begin{figure}[h]
\epsscale{1.0}\plotone{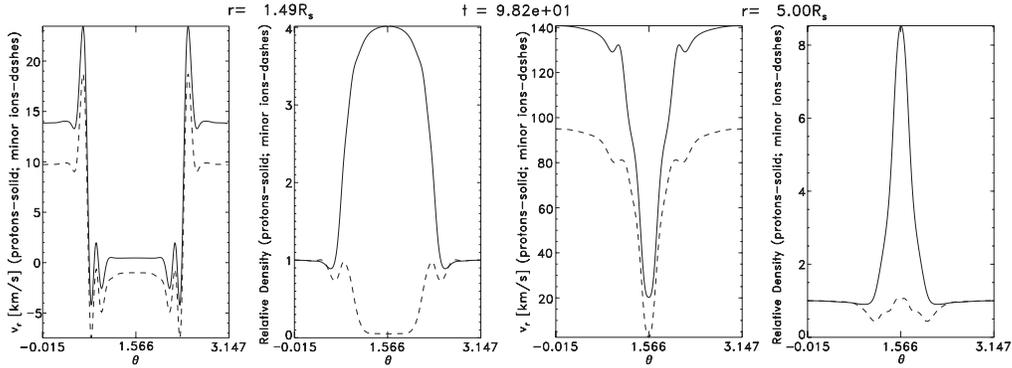} \caption{The $\theta$-
dependence at $r=1.5R_s$ (left two panels) and at $r=5R_s$ (right
two panels) for the dipolar streamer proton and $He^{++}$ outflow
speed and density.} \label{vrtheta:fig}\end{figure}

In Figures~\ref{field_3f:fig}-\ref{vrtheta:fig} we show the results
of the 3-fluid slow solar wind model. In Figure~\ref{field_3f:fig}
we show the initial dipole field in our $r-\theta$ coordinates, and
the transformation of the initially potential magnetic field to the
slow solar wind streamer with the current sheet.

In Figure~\ref{str_3f:fig} we shown the quasi-steady state solution
obtained for the slow solar wind streamer with He$^{++}$ and
O$^{5+}$ ions. The density and velocity structure of the slow solar
wind streamer is evident. The regions of highest ion density as a
function of $\theta$ occur at the interface of escaping and confined
plasma regions and at the streamer stalk, while the lowest heavy ion
density occurs at the central (confined) part of the streamer. The
He$^{++}$ ion density and velocity structure is similar to the
O$^{5+}$ ion structure. The details of the streamer cross-sectional
structure at $r=1.5R_s$ and at $r=5R_s$ for protons and He$^{++}$
ions is shown in Figure~\ref{vrtheta:fig}. The anti-correlation
between the protons and He$^{++}$ abundance and the relation to the
outflow speed is evident. The proton density dependence with height
is in qualitative agreement with Figure~\ref{Fig_vert}.

\section{Conclusions}

During the solar minimum the solar magnetic field is dominated by a
tilted dipole, and the slow solar wind is associated with a single
streamer belt, as evident from 3D reconstruction using STEREO/COR1
and NSO/GONG data. SOHO/UVCS observations of Ly-$\alpha$, O~VI, and
other minor ion emission lines in streamers provide clues for the
acceleration and heating mechanism, and require multi-fluid and
kinetic modeling in order to interpret the results.

We find that 3-fluid models of compositional structure of streamers
initiated with a dipole field reproduce the observational properties
of streamers. In particular, the global symmetry, the height
dependence of the proton density is reproduced, and the
anti-correlation with heavy ion density is recovered. We plan to
expand this study by including nonthermal sources, such as resonant
and non-resonant waves, and turbulence to study slow solar wind in
multi-ion streamers.

\acknowledgements 
This study was supported by NASA grants NNX08AV\-88G, NNX08AP88G,
and NNX08AF85G. MK would like to thank Gordon Petrie for comments
about PFSS reconstruction and Janet Luhman for PFSS code used here.


\end{document}